\documentclass[twocolumn,showpacs,preprintnumbers,amsmath,amssymb]{revtex4}
\usepackage{graphicx}% Include figure files
\usepackage{dcolumn}% Align table columns on decimal point
\usepackage{bm}% bold math
\usepackage{color}

\begin{document}

\preprint{}

\title{Spin-boson coupling in continuous-time quantum Monte Carlo}% Force line breaks with \\

\author{Junya Otsuki$^{1,2}$}
\affiliation{%
$^1$Theoretical Physics III, Center for Electronic Correlations and Magnetism,\\
Institute of Physics, University of Augsburg, D-86135 Augsburg, Germany\\
$^2$Department of Physics, Tohoku University, Sendai 980-8578, Japan
}%

\date{\today}% It is always \today, today,

\begin{abstract}
A vector bosonic field coupled to the electronic spin is treated by means of the continuous-time quantum Monte Carlo method.
In the Bose Kondo model with a sub-Ohmic density of states 
$\rho_{\rm B}(\omega) \propto \omega^{s}$ with $s=0.2$, 
two contributions to the spin susceptibility, the Curie term $T^{-1}$ and the term $T^{-s}$ due to bosonic fluctuations, are observed separately.
This result indicates the existence of a residual moment and a hidden critical behavior.
By including hybridization with itinerant electrons,
a quantum critical point is identified between this local-moment state and the Kondo singlet state. 
It is demonstrated that the energy scale of the bosonic fluctuations is not affected 
by the quantum phase transition.
\end{abstract}

% 75.20.Hr Local moment in compounds and alloys; Kondo effect, valence fluctuations, heavy fermions 
% 71.10.-w Theories and models of many-electron systems

\maketitle

\section{Introduction}
The continuous-time quantum Monte Carlo (CT-QMC) method for fermions has been developing since 2005,
as a numerical tool for correlated electron systems.\cite{Rubtsov05, Gull11}
In particular, the algorithm based on the expansion around the atomic limit (CT-HYB)\cite{Werner06,Haule07,Lauchli09} is highly effective as the impurity solver for the dynamical mean-field theory (DMFT).
The method has also been applied to variants of Kondo models,\cite{Otsuki-CTQMC, Hoshino09} where a localized spin interacts with itinerant electrons via the exchange coupling.

There is another class of impurity models which include an additional bosonic field coupled to local degrees of freedom.
The simplest one is the coupling between the electronic charge $n_f$ and a boson $\phi$ of the form $n_f \phi$.
In CT-QMC, arbitrary energy dispersion of the bosonic field is treatable, and a dynamical screening effect has been investigated.\cite{Werner07}
This algorithm can also be applied to the coupling $S_f^z \phi$ with $S_f^z$ being $z$-component of the local spin.\cite{Pixley10}
We may consider more complicated interaction including a spin flip scattering of the bosonic field, 
i.e., the coupling $\bm{S}_f \cdot \bm{\phi}$, where a vector bosonic field $\bm{\phi}$ couples to the electronic spin $\bm{S}_f$.

The coupling $\bm{S}_f \cdot \bm{\phi}$ appears when the Heisenberg interaction is treated in a ``mean-field" theory.
The boson $\bm{\phi}$ describes a time-dependent auxiliary field which mediates the effective local spin-spin interaction resulting from the intersite interaction. 
This bosonic dynamical ``bath'' is determined self-consistently, 
and thus gives descriptions of
a quantum spin glass in infinite dimensions,\cite{Bray-Moore80, Sachdev-Ye93, Grempel98, Georges00}
fluctuations around the molecular field in the (non-random) Heisenberg model,\cite{Kuramoto-Fukushima98}
and an impurity embedded in an antiferromagnet.\cite{Vojta00}
With a fermionic bath in terms of DMFT,\cite{Georges96} 
doping of the spin glass\cite{Parcollet99} and 
an extended Hubbard model with intersite interactions\cite{Smith-Si00, Haule02, Sun-Kotliar02, Rubtsov12} 
can be addressed beyond the molecular-field approximation.

These single-site theories for the Heisenberg interactions
lead to
% motivate investigations of 
the effective impurity model 
consisting of the fermionic bath $a_{\bm{k} \sigma}$ and the vector bosonic bath $b_{\bm{q} \xi}$ ($\xi=x,y,z$),
with self-consistent equations.
Solving the equations 
requires a reliable method to compute dynamical quantities of the impurity problem.
Furthermore, properties of the impurity model itself need to be understood,
since the self-consistent solution for the lattice problem inherits features of the impurity problem. 
The impurity Hamiltonian reads
\begin{align}
\label{eq:H_imp}
H 
&= \sum_{\sigma} \epsilon_{f\sigma} n_{f\sigma} 
+ U n_{f\uparrow} n_{f\downarrow}
+ \sum_{\bm{k} \sigma} \epsilon_{\bm{k}} a_{\bm{k} \sigma}^{\dag} a_{\bm{k} \sigma}
\nonumber
\\
&+ V \sum_{\sigma} ( f_{\sigma}^{\dag} a_{\sigma} + a_{\sigma}^{\dag} f_{\sigma} )
+ \sum_{\bm{q} \xi} \omega_{\bm{q} \xi} b_{\bm{q} \xi}^{\dag} b_{\bm{q} \xi}
+ \sum_{\xi} g_{\xi} S_f^{\xi} \phi^{\xi},
\end{align}
where
$a_{\sigma} = N^{-1/2} \sum_{\bm{k}} a_{\bm{k} \sigma}$,
$\phi^{\xi} = b_{\xi} + b_{\xi}^{\dag}$, and
$b_{\xi} = N^{-1/2} \sum_{\bm{q}} b_{\bm{q} \xi}$
with $N$ being the number of sites.
$n_{f\sigma} = f_{\sigma}^{\dag} f_{\sigma}$, and
$S_f^{\xi} = (1/2) \sum_{\sigma \sigma'} f_{\sigma}^{\dag} \sigma^{\xi}_{\sigma \sigma'} f_{\sigma'}$ 
with $\sigma^{\xi}$ being the Pauli matrix.
We have introduced XXZ-type anisotropy,
$g_x = g_y \equiv g_{\perp}$ and
$\omega_{\bm{q} x} = \omega_{\bm{q} y} \equiv \omega_{\bm{q} \perp}$, 
so that the formalism in this paper covers the Ising- and XY-type couplings as well.
The bosonic part in $H$ is reminiscent of the spin-boson model, which has been investigated in the context of dissipative systems.\cite{Leggett87,Bulla05}
Its SU(2) symmetric version is referred to as the Bose Kondo model,\cite{Vojta00}
and the Bose-Fermi Kondo model
with inclusion of the fermionic field.\cite{Smith-Si00}
The Hamiltonian (\ref{eq:H_imp}) describes charge fluctuations as well, and may be addressed as a Bose-Fermi Anderson model.

The essence of this model is that the fermionic field screens the localized spin, 
while the bosonic field stabilizes the moment to decouple the fermionic field.
This competition, in a certain situation, leads to a quantum phase transition between the Kondo singlet state for small $g$ and a local-moment state with a residual moment for large $g$.\cite{Si01}
Furthermore, when two or three spin directions are favored by degenerate bosonic fields,
the local-moment state may be governed by an intermediate-coupling (critical) fixed point.\cite{Vojta00}
A critical nature of this fixed point has been clarified by means of perturbative renormalization group (RG) theory.\cite{Sengupta00, Zhu-Si02, Zarand-Demler02,Vojta06}

In numerical approaches, on the other hand,
the case of Ising-type coupling (single-component bosonic field) has been extensively investigated with\cite{Glossop05} and without\cite{Bulla05, Winter09} the fermionic field either by QMC or numerical renormalization group (NRG) method.
We note that the local-moment state, in this case, is governed by a strong-coupling fixed point.
The XY-type coupling (two-component bosons) has recently been treated without fermions 
% by a variational matrix product state method.\cite{Guo12}
by using a matrix product state.\cite{Guo12}
It was found that the region of the critical phase is limited in the parameter space compared to the prediction by the RG.
This result has convinced the importance of numerical investigations.
A general situation with three-component bosonic field as well as the two-component model with the fermionic field 
have so far not been addressed by numerically reliable methods.

The purpose of this paper is twofold.
The first is to present an algorithm based on CT-QMC for solving the model (\ref{eq:H_imp}), 
which includes both the fermionic and three-component bosonic fields.
It enables us to compute static and dynamical quantities for finite temperatures, 
and could be complemental to other numerical techniques such as NRG.\cite{note-NRG, Vojta12}
Sec.~\ref{sec:ctqmc} is devoted to the explanation of the method.
Here, we restrict ourselves to $U=\infty$,
which is related to the $t$-$J$ model and the Heisenberg model 
in terms of the extended DMFT.
The second purpose of this paper is to present the first numerical results for the impurity models with the SU(2) spin-boson coupling.
We begin with a pure bosonic system without the fermionic field (Bose Kondo model) in Sec.~\ref{sec:results-BK}.
We shall demonstrate that there exists 
% instead of the critical phase, 
a localized phase in which the spin susceptibility consists of the Curie term as well as the critical term due to the bosonic fluctuations.
By including the fermionic field,
a quantum critical point is explored in Sec.~\ref{sec:results-BFA}.
We close this paper, in Sec.~\ref{sec:summary}, with a brief description of possible applications of our method.

\section{Spin-Boson Coupling in CT-QMC}
\label{sec:ctqmc}
We solve the effective impurity model (\ref{eq:H_imp}) using the hybridization-expansion solver of the CT-QMC.\cite{Werner06, Gull11}
In this section, we present how to treat the additional bosonic field in CT-QMC.

The bosonic field coupled to the electronic charge has been treated by Werner and Millis.\cite{Werner07}
In this method, the electron-phonon coupling is eliminated by the so-called Lang-Firsov transformation,
and it makes the computation efficient. 
This manipulation can also be applied to the coupling between $S_f^z$ and bosons.\cite{Pixley10}
In the case of the exchange coupling,
however, 
we cannot eliminate it by this transformation,
% simultaneously,
since three components of the spin operators $\bm{S}_f$
% $S_f^x$, $S_f^y$, and $S_f^z$, 
do not commute with each other.
Only one component can be eliminated among three.
Hence, 
we treat the other two by a stochastic method.
% Namely, the spin-flip scattering of the bosonic fields as well as the hybridization with the fermionic bath are treated by a Monte Carlo method.
Namely, we perform expansions with respect to the spin-flip scattering as well as the hybridization, and sum up the series by a Monte Carlo sampling.

Before proceeding to the formulation,
we define the propagators for the fermionic field (hybridization function) and 
the bosonic field (effective interaction) as follows: 
\begin{align}
\label{eq:bath-f}
\Delta(i\omega_n) 
&= 
V^2 G_0(i\omega_n) 
=\frac{V^2}{N} \sum_{\bm{k}} \frac{1}{i\omega_n - \epsilon_{\bm{k}}},
\\
\label{eq:bath-b}
{\cal J}_{\gamma}(i\nu_n) 
&= 
-g_{\gamma}^2 D_{\gamma 0} (i\nu_n)
= \frac{g_{\gamma}^2}{N} \sum_{\bm{q}} \frac{2\omega_{\bm{q} \gamma}}{\nu_n^2 + \omega_{\bm{q} \gamma}^2},
\end{align}
where 
$\omega_n=(2n+1)\pi T$ and $\nu_n=2n \pi T$ are the fermionic and bosonic Matsubara frequencies
and $\gamma =z, \perp$.
The latter quantity describes the effective interaction 
$-\bm{S}_f(\tau) \cdot {\cal J}(\tau-\tau') \bm{S}_f(\tau')$
mediated by the bosonic field.

\subsection{Canonical transformation}

We first eliminate the coupling between $S_f^z$ and bosons.
Following Ref.~\cite{Werner07},
we perform a canonical transformation 
$\tilde{H} = e^{\cal S} H e^{-{\cal S}}$ with 
${\cal S}= N^{-1/2} \sum_{\bm{q}} (g_z/\omega_{\bm{q}z}) (b_{\bm{q}z}^{\dag} - b_{\bm{q}z}) S_f^z$,
which shifts the $z$-coordinate of the oscillation to eliminate the term $g_z S_f^z \phi^z$.
The transformed Hamiltonian $\tilde{H}$ is given by
\begin{align}
\tilde{H}
&= 
\sum_{\sigma} \tilde{\epsilon}_{f\sigma} n_{f\sigma} 
+ \tilde{U} n_{f\uparrow} n_{f\downarrow}
+ \sum_{\bm{k} \sigma} \epsilon_{\bm{k}} a_{\bm{k} \sigma}^{\dag} a_{\bm{k} \sigma}
\nonumber \\
&+ V \sum_{\sigma} ( \tilde{f}_{\sigma}^{\dag} a_{\sigma} + a_{\sigma}^{\dag} \tilde{f}_{\sigma} )
+ \sum_{\bm{q} \xi} \omega_{\bm{q} \xi} b_{\bm{q} \xi}^{\dag} b_{\bm{q} \xi}
\nonumber \\
&+ \frac{g_{\perp}}{\sqrt{2}} (\tilde{S}_f^+ \phi^- + \tilde{S}_f^- \phi^+),
\end{align}
where
$\phi^{\pm} = (\phi^x \pm i\phi^y)/\sqrt{2}$.
% $[\phi^{+}, \phi^{-}]=$.
The local parameters are renormalized to
$\tilde{\epsilon}_{f\sigma} = \epsilon_{f\sigma} - N^{-1} \sum_{\bm{q}} g_z^2 / (4\omega_{\bm{q}z})$
and
$\tilde{U} = U + N^{-1} \sum_{\bm{q}} g_z^2 / (2\omega_{\bm{q}z})$.
The operators for the local electron are transformed to
\begin{align}
\tilde{f}_{\sigma} = e^{-\sigma A/2} f_{\sigma},
\quad
\tilde{f}_{\sigma}^{\dag} = e^{\sigma A/2} f_{\sigma}^{\dag},
\quad
\tilde{S}_f^{\pm} = e^{\pm A} S_f^{\pm},
\label{eq:phase}
\end{align}
where
$S_f^{\pm} = S_f^x \pm iS_f^y$ and
$
A = N^{-1/2} \sum_{\bm{q}} (g_z/\omega_{\bm{q}z})
(b_{\bm{q}z}^{\dag} - b_{\bm{q}z}).
$
In Eq.~(\ref{eq:phase}), the factor $e^{A /2}$ is associated with the change in the quantum number of $S_f^z$.

\subsection{Partition function}
\begin{figure}[tb]
	\begin{center}
	\includegraphics[width=7cm]{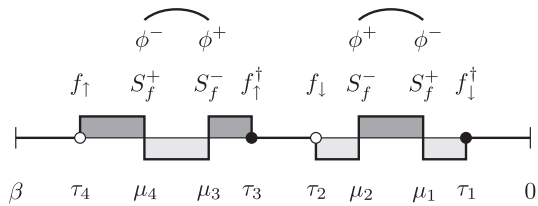}
	\end{center}
	\caption{An example of the Monte Carlo configuration of order $k=2$ and $l=2$. The dark and light shaded area indicate the spin-up and -down states, respectively. The curved lines express the bosonic Green function.}
	\label{fig:config}
\end{figure}

With the transformed Hamiltonian $\tilde{H}$, 
we expand the partition function $Z$ with respect to $V$ and $g_{\perp}$ as follows:
\begin{align}
\label{eq:part_func}
\frac{Z}{Z_0} 
= \sum_{k=0}^{\infty} \sum_{l=0}^{\infty} 
\int d\bm{\tau} \int d\bm{\mu} 
W(\bm{\tau}, \bm{\mu}),
\end{align}
where the subscript 0 denotes a quantity for 
$V=g_{\perp}=0$.
The integrand $W(\bm{\tau}, \bm{\mu})$ describes the contribution of order $V^{2k} g_{\perp}^{2l}$.
The variables
$\bm{\tau} =(\tau_1, \cdots, \tau_{2k})$ and
$\bm{\mu} = (\mu_1, \cdots, \mu_{2l})$
denote sets of imaginary times at which the hybridization and spin exchange events occur, respectively. 
The integrals are taken over the range $\beta > \tau_{2k} > \cdots > \tau_1 \geq 0$,
and the same for $\bm{\mu}$.
Figure \ref{fig:config} shows an example of the configuration. 
The doubly occupied state is excluded in this figure, since we consider the limit $U=\infty$ in the next subsection.
We note that the formulae in this subsection are valid also for $U<\infty$.
In the simulation, the summations over $k$ and $l$ as well as the integrals over $\bm{\tau}$ and $\bm{\mu}$ are to be evaluated via an importance sampling.

The weight $W(\bm{\tau}, \bm{\mu})$ is decoupled into four contributions
according to types of operators:
\begin{align}
W(\bm{\tau}, \bm{\mu}) = 
\tilde{W}_{\rm loc}(\bm{\tau}, \bm{\mu}) 
W_{\rm hyb}(\bm{\tau}) W_{\perp}(\bm{\mu}) W_z(\bm{\tau}, \bm{\mu}). 
\end{align}
The first two are the contributions from the Anderson model:\cite{Werner06}
the local contribution $\tilde{W}_{\rm loc}$ 
% denotes the trace over the $f$-state and 
is simply given by the Boltzmann factor with
the renormalized parameters, $\tilde{\epsilon}_{f\sigma}$ and $\tilde{U}$,
and $W_{\rm hyb}$ denotes the trace over the fermionic field, 
which is expressed by the determinant of a $k \times k$ matrix 
consisting of $\Delta(\tau)$ in Eq.~(\ref{eq:bath-f}).
In the following, we explain the bosonic contributions in turn.

The third factor $W_{\perp}(\bm{\mu})$ incorporates the $xy$-component of the bosonic operators appearing in the series expansion with respect to $g_{\perp}$:
\begin{align}
W_{\perp} (\bm{\mu})
= \frac{g_{\perp}^{2l}}{2^l}
\langle \phi^{\eta_{2l}}(\mu_{2l}) \cdots \phi^{\eta_{1}}(\mu_1) \rangle_0,
\end{align}
where $\phi^{\eta}$ denotes either $\phi^+$ or $\phi^-$.
The numbers of $\phi^+$ and $\phi^-$ must be the same, since we have the relation
$\langle \phi^{\pm}(\mu_i) \phi^{\pm}(\mu_j) \rangle_0 =0$.
The thermal average in $W_{\perp}$ is decomposed by Wick's theorem, and is represented by the permanent of an $l \times l$ matrix consisting of
$D_{\perp 0}(\mu_i-\mu_j)=-\langle T_{\tau} \phi^{+}(\mu_i) \phi^{-}(\mu_j) \rangle_0$.\cite{footnote-pm}
However, since there is no efficient algorithm for computing the permanent,
we evaluate it by a stochastic sampling.\cite{Anders11} 
Namely, we express $W_{\perp}$ as
\begin{align}
\label{eq:sum_alpha}
W_{\perp} (\bm{\mu}) = \sum_{\alpha} W_{\perp} (\bm{\mu}; \alpha),
\end{align}
with $\alpha$ denoting one of terms in the permanent, 
and the summation is to be evaluated stochastically.
In Fig. \ref{fig:config}, the configuration $\alpha$ is represented by curved lines.

The last contribution $W_z(\bm{\tau}, \bm{\mu})$ is due to the $z$-component of the bosonic field, which is now expressed as the phase factors in Eq.~(\ref{eq:phase}). 
The explicit expression is given by
\begin{align}
W_z (\bm{\tau}, \bm{\mu}) = \langle e^{s_{2m} A(t_{2m})} \cdots e^{s_1 A(t_1)} \rangle_0,
\end{align}
where
$\{ t_i \}$ is composed of $\bm{\tau}$ and $\bm{\mu}$ in ascending order and
$m=k+l$.
$A(t) = N^{-1/2} \sum_{\bm{q}} (g_z / \omega_{\bm{q}z}) 
( e^{\omega_{\bm{q}z} t} b_{\bm{q}z}^{\dag} - e^{-\omega_{\bm{q}z} t} b_{\bm{q}z} )$.
The factor $s_i$ takes 
$\sigma/2$ for $f_{\sigma}^{\dag}$,
$-\sigma/2$ for $f_{\sigma}$,
and 
$\pm 1$ for $S_f^{\pm}$.
Using the condition $\sum_i s_i = 0$,
the thermal average can be evaluated analytically to give\cite{Werner07}
\begin{align}
&W_z(\bm{\tau}, \bm{\mu})
= \exp \left[ \sum_{2m \geq j> i \geq 1} s_i s_j K(t_j - t_i) \right], \\
&K(\tau) = -\frac{1}{N} \sum_{\bm{q}} \frac{g_z^2}{\omega_{\bm{q}z}^2} [B(\omega_{\bm{q}z}, \tau) - B(\omega_{\bm{q}z}, 0)],
% &B(\omega, \tau) = \frac{\cosh (\beta/2 - \tau) \omega}{\sinh (\beta \omega/2)}$.
\label{eq:K_tau}
\end{align}
where
$B(\omega, \tau) = \cosh [(\beta/2 - \tau) \omega] /\sinh (\beta \omega/2)$.

So far, we have used $\omega_{\bm{q}z}$ explicitly,
but actually the dynamics of the bosonic field enters only through the function
${\cal J}_{\gamma}(i \nu_n)$
defined in Eq.~(\ref{eq:bath-b}).
It is therefore convenient to express the summations over $\bm{q}$ 
% in the above formulae 
in terms of ${\cal J}_{\gamma}(i\nu_n)$.
The renormalized parameters are rewritten as
$\tilde{\epsilon}_{f\sigma} = \epsilon_{f\sigma} - {\cal J}_z(0)/8$ and
$\tilde{U} = U + {\cal J}_z(0)/4$.
The function $K(\tau)$ in Eq.~(\ref{eq:K_tau}) is rewritten as
\begin{align}
K(\tau) 
% &= T \sum_n g_z^2 D_0(i\nu_n) \frac{(-1)^n}{(i\nu_n)^2} 
% [ \cos (\beta /2 - \tau) \nu_n - \cos (\beta \nu_n /2)] \\
= 
{\cal J}_z(0) \frac{\tau (\beta - \tau)}{2\beta} 
- \sum_{n \neq 0} {\cal J}_z(i\nu_n) \frac{1- \cos \tau \nu_n}{\beta \nu_n^2}.
\end{align}

\subsection{Monte Carlo procedure}
\label{sec:mc}
We perform stochastic samplings of $\bm{\tau}$ and $\bm{\mu}$ in Eq.~(\ref{eq:part_func})
and $\alpha$ in Eq.~(\ref{eq:sum_alpha}).
They respectively correspond to the $V$-expansion, $g_{\perp}$-expansion and
the Wick's theorem for the bosonic field.
Since the Hamiltonian with $V=g_{\perp}=0$ conserves the quantum number of $S_f^z$, 
we can treat $\tilde{W}_{\rm loc}$ by the ``segment picture" of CT-HYB.\cite{Werner06, Gull11}
Hence for the $V$-expansion, the update procedure in the Anderson model can be used.\cite{Werner07}
Hereafter, we consider the limit $U=\infty$, which can be implemented by excluding the doubly occupied state in the configuration.

\begin{figure}[tb]
	\begin{center}
	\includegraphics[width=7cm]{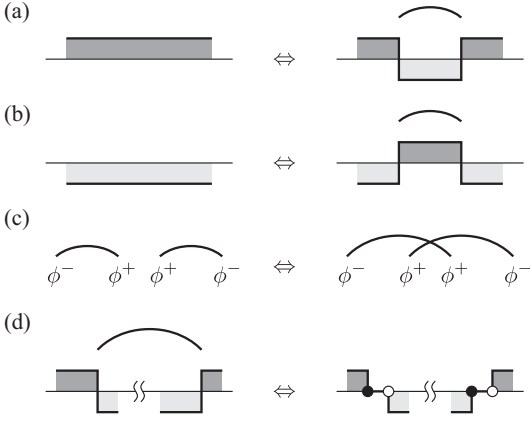}
	\end{center}
	\caption{Update processes necessary to evaluate the spin-boson coupling.}
	\label{fig:update}
\end{figure}

In addition to the updates in CT-HYB, 
we perform the following updates to sum up $g_{\perp}$-terms: 
\begin{enumerate}
% \item Addition/Removal of ($f_{\sigma}, f_{\sigma}^{\dag}$) on empty state
% \item Addition/Removal of ($f_{\sigma}^{\dag}, f_{\sigma}$) on $\sigma$-state
\item[(a)] Insertion/Removal of $S_f^+(\mu+\ell) S_f^-(\mu)$ on $\uparrow$-state.
\item[(b)] Insertion/Removal of $S_f^-(\mu+\ell) S_f^+(\mu)$ on $\downarrow$-state.
\item[(c)] % Exchange of links of bosonic Green function.
Change of the configuration $\alpha$.
\item[(d)] Replacing $S_f^+(\mu_i')$ and $S_f^-(\mu_i)$ 
with $f_{\uparrow}^{\dag}(\mu_i'+\ell') f_{\downarrow}(\mu_i')$ 
and $f_{\downarrow}^{\dag}(\mu_i+\ell) f_{\uparrow}(\mu_i)$,
and vice versa.
\end{enumerate}
These updates are expressed diagrammatically in Fig. \ref{fig:update}.
The updates (a) and (b) change the expansion order of $g_{\perp}$ by 2.
In (c), we exchange two links of the bosonic Green functions.
The ergodicity is in principle satisfied only by (a)--(c).
However, a part of the configuration may freeze in practice, 
when the expansion orders for $g_{\perp}$ and $V$ are considerably different from each other, say, when $g_{\perp}$ is much smaller than $V$.
The freezing happens because a pair of spin operators between which hybridization operators are located cannot be removed by the updates (a) and (b).
% The freezing happens because the spin operators are separated from each other by hybridization operators located in between.
% In such a situation, the spin operators cannot be removed by the updates (a) and (b).
This problem can be resolved by introducing the update (d), 
which replaces two spin operators separated in the time ordering 
with single-particle operators.
[This update is important, for example, in the parameter range $0<g \lesssim 0.12$ in Fig.~\ref{fig:renorm}.]

We first consider the update (a).
In the insertion process, we choose two imaginary times randomly in the same way as the ``segment algorithm"\cite{Werner06, Gull11}: 
$\mu$ is first chosen from the full range $[0:\beta)$ and then 
the length $\ell$ is chosen from the restricted range $(0:\ell_{\rm max})$ 
% so that the resultant configuration is allowed.
so that the operator $S_f^+$ does not pass the next operators.
In the removal process, 
we choose one pair from $(l+1)$ pairs of the spin operators which are connected by the bosonic line,
and try the update if it is allowed, 
i.e., if no operator exists between them.
From the detailed balance condition, 
the update probability $R$ 
is given by
\begin{align}
% \frac{p(l \to l+1)}{p(l+1 \to l)}
% \frac{p(\bm{\mu} \to \bm{\mu}^+)}{p(\bm{\mu}^+ \to \bm{\mu})}
R(\bm{\mu} \to \bm{\mu}^+)
= \frac{\beta \ell_{\rm max}}{l+1}
% \frac{g_{\perp}^2}{2} D_0(\mu - \mu') 
\frac{{\cal J}_{\perp}(-\ell)}{2} 
% e^{(\tau'-\tau)(\epsilon_{f\uparrow} - \epsilon_{f\downarrow})} 
\frac{\tilde{W}_{\rm loc} (\bm{\tau}, \bm{\mu}^+)}{\tilde{W}_{\rm loc} (\bm{\tau}, \bm{\mu})}
\frac{W_z (\bm{\tau}, \bm{\mu}^+)}{W_z (\bm{\tau}, \bm{\mu})},
% \frac{\tilde{W}_{\rm loc}^{(m+1)}}{\tilde{W}_{\rm loc}^{(m)}}
% \frac{W_z^{(m+1)}}{W_z^{(m)}}.
\label{eq:prob_ins}
\end{align}
where  $\bm{\mu}$ and $\bm{\mu}^{+}$ denote 
the configurations of order $g_{\perp}^{2l}$ and $g_{\perp}^{2(l+1)}$, respectively.
The expression for the update (b) is given in a similar manner.

The update probability for (c) comes only from $W_{\perp}$.
Suppose that 
$(\mu_i, \mu_i')$ and $(\mu_j, \mu_j')$ denote pairs of imaginary times connected by the bosonic Green function in the original configuration $\alpha$.
Then, the update probability $R$ for exchanging the links is given by
\begin{align}
% \frac{p(\alpha \to \alpha')}{p(\alpha' \to \alpha)}
R(\alpha \to \alpha')
% = \frac{W_{\perp}(\bm{\mu}; \alpha')}{W_{\perp}(\bm{\mu}; \alpha)}.
= \frac{{\cal J}_{\perp}(\mu_i - \mu_j') {\cal J}_{\perp}(\mu_j - \mu_i')}{{\cal J}_{\perp}(\mu_i - \mu_i') {\cal J}_{\perp}(\mu_j - \mu_j')}.
\label{eq:prob_exc}
\end{align}

Finally, we consider the update (d).
We first choose a pair of spin operators connected by the bosonic line, as in the removal process of the update (a).
They are to be replaced by $f_{\downarrow}$ and $f_{\uparrow}$, respectively.
Simultaneously, the operator $f^{\dag}_{\uparrow}$ ($f^{\dag}_{\downarrow}$) is placed next to $f_{\downarrow}$ ($f_{\uparrow}$).
Here, the length $\ell$ ($\ell'$) of the empty state is chosen from the range up to $\ell_{\rm max}$ ($\ell'_{\rm max}$) so that the resultant configuration is allowed.
In the opposite process,
we choose the operators $f_{\downarrow}$ and $f_{\uparrow}$ 
from $(k_{\downarrow}+1)$ and $(k_{\uparrow}+1)$ randomly,
where $k_{\sigma}$ denotes the hybridization-expansion order for spin $\sigma$.
The update probability $R$ 
% from the original configuration $(\bm{\tau}, \bm{\mu})$ 
% of order $V^{2k} g_{\perp}^{2l}$ 
% to the new configuration $(\bm{\tau}^{++}, \bm{\mu}^{-})$ 
% of order $V^{2(k+2)} g_{\perp}^{2(l-1)}$ 
is given by
\begin{align}
R&(\bm{\tau}, \bm{\mu} \rightarrow \bm{\tau}^{++}, \bm{\mu}^-) 
= \frac{l \ell_{\rm max} \ell'_{\rm max}}{(k_{\uparrow}+1) (k_{\downarrow}+1)}
\frac{2}{{\cal J}_{\perp}(\mu_i - \mu_i')}
\nonumber \\
&\times
% \left( -\frac{W_{\rm hyb} (\bm{\tau}^{++})}{W_{\rm hyb} (\bm{\tau})} \right)
\frac{W_{\rm hyb} (\bm{\tau}^{++})}{W_{\rm hyb} (\bm{\tau})}
\frac{\tilde{W}_{\rm loc} (\bm{\tau}^{++}, \bm{\mu}^{-})}{\tilde{W}_{\rm loc}(\bm{\tau}, \bm{\mu})}
\frac{W_{z} (\bm{\tau}^{++}, \bm{\mu}^{-})}{W_{z} (\bm{\tau}, \bm{\mu})},
\end{align}
where $\bm{\tau}^{++}$ and $\bm{\mu}^{-}$ denote the new configuration 
of order $V^{2(k+2)} g_{\perp}^{2(l-1)}$.

We have confirmed, in the simulation, 
that all the update probabilities presented above
are always positive
and therefore, the simulation does not suffer from the sign problem.

\subsection{Spin susceptibility}
We define the spin susceptibilities by
$\chi_{zz}(\tau) = \langle S_f^z (\tau) S_f^z \rangle$ and
% $\chi_{\pm \mp}(\tau) = \langle S_f^{\pm}(\tau) S_f^{\mp} \rangle /2$.
$\chi_{+-}(\tau) = \langle S_f^{+}(\tau) S_f^{-} \rangle /2$.
In the isotropic system, we have $\chi_{zz}(\tau)=\chi_{+-}(\tau)$.
We can evaluate $\chi_{zz}(\tau)$ from the configuration of the $f$-operators as in the ``segment algorithm"\cite{Werner06, Gull11}. 
On the other hand, $\chi_{+-}(\tau)$ can be evaluated by
\begin{align}
\chi_{+-} (\tau) 
% = -\frac{T}{g_{\perp}^2} \left< \sum_{i=1}^{l} \frac{\delta(\tau, \mu_i - \mu_i')}{D_0(\mu_i - \mu_i')} \right>_{\rm MC},
= T \left< \sum_{i=1}^{l} \frac{\delta(\tau, \mu_i' - \mu_i)}{{\cal J}_{\perp}(\mu_i - \mu_i')} \right>_{\rm MC},
\label{eq:chi_pm}
\end{align}
where $\mu_i$ and $\mu_i'$ denote the imaginary times for $S_f^-$ and $S_f^+$ which are connected by the bosonic Green function,
% where $\mu_i$ denotes the imaginary time for $S_f^+$ and $\mu_i'$ for $S_f^-$ which are ``connected" to $\mu_i$ in the configuration $\alpha$.
% where the summation of $i$ is taken for all the bosonic Green functions in the configuration $\alpha$, 
and MC means average over Monte Carlo configuration.
The function $\delta(\tau, \mu)$ is defined by
\begin{align}
\delta(\tau, \mu) = \left\{
	\begin{array}{ll}
	\delta(\tau-\mu) & (\mu>0) \\
	\delta(\tau-\mu-\beta) & (\mu<0)
	\end{array}
\right.,
\end{align}
and $\chi_{+-}(\tau)$ is sampled in the range $0<\tau<\beta$.
The end points are evaluated accurately from the occupation number using the relations
$\chi_{+-}(+0)= \langle n_{f\uparrow} \rangle/2$ and
$\chi_{+-}(\beta -0) = \langle n_{f\downarrow} \rangle/2$.
Equation~(\ref{eq:chi_pm}) follows from the fact that
${\cal J}_{\perp}$ describes the retarded interaction between the local spin 
so that it may be regarded as a source field for the susceptibility. 

The susceptibilities $\chi_{zz}(\tau)$ and $\chi_{+-}(\tau)$ can also be computed using the matrix $M_{\sigma}$ which is kept in the simulation to evaluate the determinant in $W_{\rm hyb}$.\cite{Rubtsov05, Werner06, Gull11}
Although this way is not efficient compared to the method presented above, 
we can use it for a check of the algorithm and a code.
Another consistency check is $\chi_{zz}(\tau)=\chi_{+-}(\tau)$ in isotropic parameters,
since this condition is not trivial in the present algorithm,
which treats $g_z$ and $g_{\perp}$ in different ways. 
We have confirmed that our results satisfy this condition.

\section{Pure Bosonic System}
\label{sec:results-BK}
In this section, we present numerical results for the pure bosonic system,
i.e., the limit $V=0$ and $U=-\epsilon_f=\infty$.
The charge fluctuation is absent in this limit 
so that the local electron is reduced to a localized spin $\bm{S}$.
In the present algorithm, the elimination of the charge fluctuation can be easily implemented by restricting the updates to (a)--(c) in Sec.~\ref{sec:mc}. 
The corresponding Hamiltonian with the SU(2) symmetry is written as
\begin{align}
H_{\rm BK} = 
 \sum_{\bm{q}} \omega_{\bm{q}} \bm{b}_{\bm{q}}^{\dag} \cdot \bm{b}_{\bm{q}}
% + h S^z_f 
% + g S^z_f \phi^z
% + \frac{g_{\perp}}{\sqrt{2}} ( S_f^{+} \phi^{-} + S_f^{-} \phi^{+}),
+ g \bm{S} \cdot \bm{\phi}.
\end{align}
This model is referred to as the Bose Kondo model or the SU(2) spin-boson model.\cite{Vojta00, Vojta06}
% which has been investigated in the context of an impurity embedded in an antiferromagnet.\cite{Vojta00, Vojta06}

The bosonic field is characterized by the density of states
$\rho_{\rm B}(\omega) = N^{-1} \sum_{\bm{q}} \delta(\omega - \omega_{\bm{q}})$.
We use a function
$\rho_{\rm B}(\omega) \propto \omega^{s}$ with a cut-off energy $\omega_{\rm c}$.
The sum-rule of the density of states, 
$\int_0^{\infty} d\omega \rho_{\rm B}(\omega)=1$, 
determines the factor to yield the explicit form
\begin{align}
\rho_{\rm B}(\omega)
= (s+1) \omega^{s} \omega_{\rm c}^{-s-1}
\theta(\omega_{\rm c} - \omega).
\label{eq:dos_B}
\end{align}
We take $\omega_{\rm c}=1$ as the unit of energy.

According to the RG analysis,\cite{Vojta00, Zhu-Si02, Zarand-Demler02, Vojta06}
this model has an intermediate-coupling fixed point (critical phase)
for $0<s<1$.
At this fixed point, the susceptibility shows the long-time behavior $\chi(\tau) \sim \tau^{1-s}$,
which indicates the static susceptibility of the form $\chi \sim T^{-s}$.
On the other hand, recent
numerical calculations for the XY-type coupling revealed
that the region $0<s<s^{*}$ ($s^{*} = 0.75$ in the limit $g\to0$) is actually 
a localized phase which does not show the critical behavior.\cite{Guo12}
Hence, this localized phase is also expected for the SU(2) coupling with $s$ close to 0.
% For $s\geq 1$, on the other hand, the bosonic coupling is irrelevant.
In the following, we investigate $s=0.2$ in detail and shall demonstrate that it indeed belongs to the localized phase.

\begin{figure}[tb]
	\begin{center}
	\includegraphics[width=\linewidth]{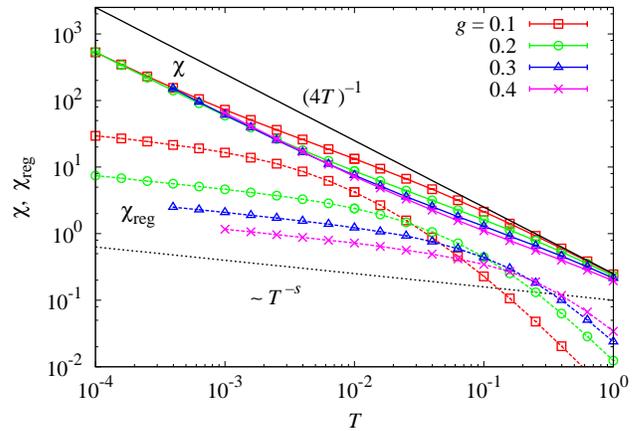}
	\end{center}
	\caption{(Color online) Temperature dependences of the static susceptibility $\chi(0)$ in the pure bosonic system with $s=0.2$ (solid lines). 
	The regular part $\chi_{\rm reg}(0)$ defined in Eq.~(\ref{eq:chi_reg}) is also plotted (dashed lines).}
	\label{fig:suscep-BK}
\end{figure}

Fig.~\ref{fig:suscep-BK} shows temperature dependences of the static spin susceptibility for $s=0.2$.
% We can clearly see the Curie behavior $\chi \propto T^{-1}$ at low temperatures 
% indicating the existence of a residual moment.
It turns out that the low-temperature susceptibility follows the Curie law $\chi \propto T^{-1}$,
indicating the existence of a residual moment.
This result demonstrates that $s=0.2$ is not in the critical phase but in the localized phase.
Nevertheless, the critical term $T^{-s}$ originating from the bosonic fluctuations still exists behind the Curie term.
To see this,
we define a regular part of the susceptibility, $\chi_{\rm reg}(z)$, by an analytical continuation of $\chi(z=i\nu_n)$ with $\nu_n>0$.
Using $\chi_{\rm reg}(z)$,
the susceptibility is written as
\begin{align}
\chi(i\nu_n) = \delta_{n0} M/4T  + \chi_{\rm reg}(i\nu_n).
\label{eq:chi_reg}
\end{align}
We note that the effective moment $M$ may depend on temperature.
The full-moment corresponds to $M=1$.
We evaluate $\chi_{\rm reg}(0)$ by an extrapolation from $\chi(i\nu_1)$, $\chi(i\nu_2)$ and $\chi(i\nu_3)$ with a quadratic function.
We have confirmed that the choice of the functional form in the extrapolation does not affect the low-temperature behavior.\cite{footnote-extrap}
The result is shown in Fig.~\ref{fig:suscep-BK}.
We clearly see the power-law behavior
$\chi_{\rm reg}(0) \propto T^{-s}$ at low temperatures.
Consequently, the low-temperature static susceptibility can be expressed in terms of two diverging terms
\begin{align}
% \chi(0) \simeq M/4T + A/T^{s}.
\chi(0) \simeq M_0/4T + 1/[4T^{s} (T_{\rm B})^{1-s}],
\label{eq:chi-BK}
\end{align}
where $M_0=\lim_{T\to 0}M$ is the residual moment
and we have introduced a characteristic energy scale $T_{\rm B}$ of the bosonic fluctuations.
\begin{figure}[tb]
	\begin{center}
	\includegraphics[width=\linewidth]{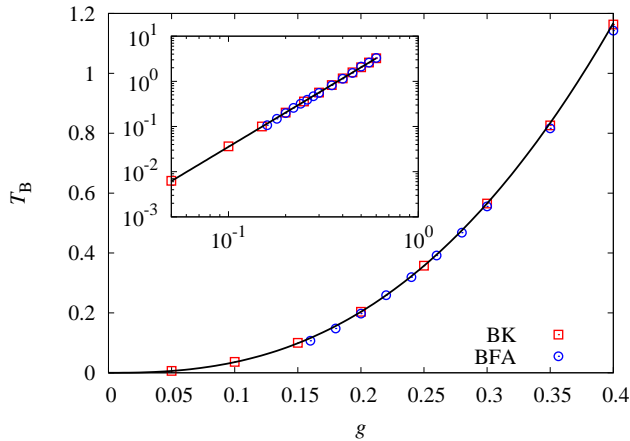}
	\end{center}
	\caption{(Color online) The energy scale $T_{\rm B}$ of the bosonic fluctuations in the pure bosonic system with $s=0.2$ (denoted by BK). A result for the Bose-Fermi Anderson model is also plotted (denoted by BFA, see Fig.~\ref{fig:suscep-BFA} for parameters).
	The solid line shows the function $T_{\rm B} \propto g^{\alpha}$ fitted to the BK data.
	%and $\alpha \simeq 2.52$ has been obtained.
	}
	\label{fig:tb}
\end{figure}
It turns out that $T_{\rm B}$ exhibits 
a power-law behavior $T_{\rm B} \propto g^{\alpha}$ as shown in Fig.~\ref{fig:tb}.
The exponent $\alpha$ is obtained as 
$\alpha \simeq 2.52$ with the error 0.01.
The residual moment $M_0$ weakly depends on $g$.
[The figure is presented in the next section (Fig.~\ref{fig:renorm}) together with results for the Bose-Fermi Anderson model.]

\section{Fermionic and Bosonic fields}
\label{sec:results-BFA}

We proceed to the system with both the bosonic and fermionic fields.
Due to the hybridization with the itinerant electrons, the Kondo fixed point with decoupled bosonic field emerges in addition to those in the pure bosonic system. 
According to the RG analysis 
for the Bose-Fermi Kondo model,
a quantum critical point characterized by $\chi \sim T^{-s}$ exists between the Kondo phase and the critical bosonic phase 
% where the fermionic bath is decoupled, 
for $0<s<1$.\cite{Zhu-Si02, Zarand-Demler02, Vojta06}
However, we should note that this result may not apply to the region away from $s=1$, since this region is not actually in the critical phase as demonstrated for $s=0.2$ in the previous section.
% However, it should be noted that the region away from $s=1$ is not actually in the critical phase as demonstrated for $s=0.2$ in the previous section, and therefore this result may not be applicable to this region.
In the following, we explore a quantum phase transition between the Kondo singlet state and the (non-critical) local-moment state.

We use the same condition for the bosonic field, $s=0.2$, as in the previous section.
For the fermionic density of states, 
$\rho_{\rm F}(\omega)= N^{-1} \sum_{\bm{k}} \delta(\omega-\epsilon_{\bm{k}})$, 
on the other hand,
we use a rectangular model with a cut-off energy $D$
\begin{align}
\rho_{\rm F}(\omega) = (1/2D) \theta(D - |\omega|).
\end{align}
We vary $g_z = g_{\perp} \equiv g$,
fixing $V^2=0.1$, $\epsilon_{f\sigma}=-0.2$, $U=\infty$ and 
$\omega_{\rm c}=D=1$.
The Kondo temperature $T_{\rm K}$ is estimated to be 
$T_{\rm K} \sim 0.1$ for $g =0$.

\begin{figure}[tb]
	\begin{center}
	\includegraphics[width=\linewidth]{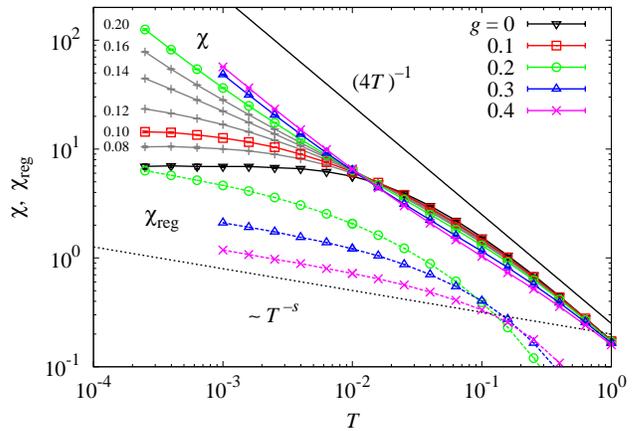}
	\end{center}
	\caption{(Color online) Temperature dependences of the static susceptibility $\chi(0)$ in the Bose-Fermi Anderson model with $V^2=0.1$, $\epsilon_{f\sigma}=-0.2$, $U=\infty$ and $s=0.2$ (solid lines).
	The dashed lines show the regular part $\chi_{\rm reg}(0)$ defined in Eq.~(\ref{eq:chi_reg}).}
	\label{fig:suscep-BFA}
\end{figure}

We show temperature dependence of the spin susceptibility in Fig.~\ref{fig:suscep-BFA}.
Difference with the pure bosonic system is the paramagnetic behavior in the small-$g$ region, $g \lesssim 0.10$. 
This indicates the spin fluctuations in the Kondo singlet state given by\cite{Hewson}
\begin{align}
	\chi(0)=1/4T_{\rm F},
\label{eq:chi-Kondo}
\end{align}
with $T_{\rm F}$ being the energy scale of low-energy excitations.
The low-temperature susceptibility increases against $g$, indicating a reduction of $T_{\rm F}$.
To quantify the local Fermi-liquid state, 
we evaluate the renormalization factor $z$ defined by
$z=[1-{\rm Im}\Sigma_f(i\omega_0)/\omega_0]^{-1}$.
\begin{figure}[tb]
	\begin{center}
	\includegraphics[width=\linewidth]{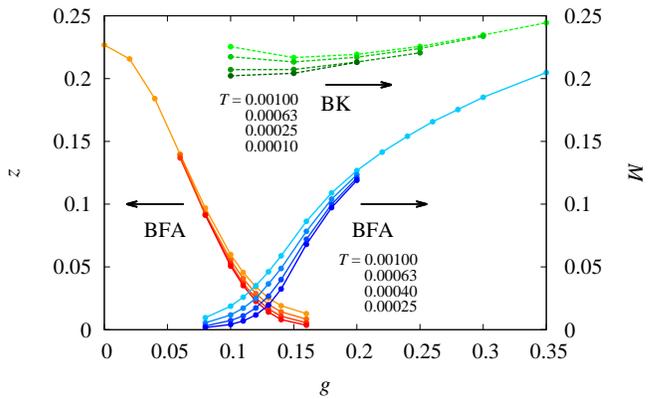}
	\end{center}
	\caption{(Color online) The renormalization factor $z$ and the effective moment $M$ for the same parameters as in Fig.~\ref{fig:suscep-BFA}. Results at four different temperatures are plotted. The dashed line (denoted by BK) is the result for the pure bosonic system.}
	\label{fig:renorm}
\end{figure}
The result is plotted in Fig.~\ref{fig:renorm} for several values of $T$. 
We can see the reduction of $z$ with increasing $g$, and
it is estimated as $z \leq 0.023$ (0.008) at $g = 0.12$ (0.14).
However, 
% since $z$ is not converged around $g= 0.12$ in this temperature range, 
since $z$ is not yet converged for $g \gtrsim 0.10$ in this temperature range, 
we cannot identify the quantum critical point from these data.

In the large-$g$ region, $g \gtrsim 0.14$, on the other hand, $\chi$ shows the Curie behavior
$\chi \propto T^{-1}$.
As in the pure bosonic system,
we evaluate the regular part $\chi_{\rm reg}(0)$ defined in Eq.~(\ref{eq:chi_reg}), which expresses the contribution 
after subtraction of the Curie term.
It turns out from Fig.~\ref{fig:suscep-BFA} that $\chi_{\rm reg}(0)$ shows the power-law behavior $T^{-s}$ as in Fig.~\ref{fig:suscep-BK}.
A remarkable point is that the energy scale $T_{\rm B}$ of the bosonic fluctuation 
% defined in Eq.~(\ref{eq:chi-BK}), 
is not affected by the hybridization as shown in Fig.~\ref{fig:tb}.
Hence, the difference to the pure bosonic system in the local-moment regime comes from the Curie term.
To see this, we evaluate the effective moment $M$ by subtracting $\chi_{\rm reg}$ from $\chi$ in Eq.~(\ref{eq:chi_reg}), 
and plot it as a function of $g$ in Fig.~\ref{fig:renorm}.
It turns out that $M$ is strongly suppressed compared to that in the pure bosonic system 
below $g \simeq 0.20$.

\begin{figure}[tb]
	\begin{center}
	\includegraphics[width=0.9\linewidth]{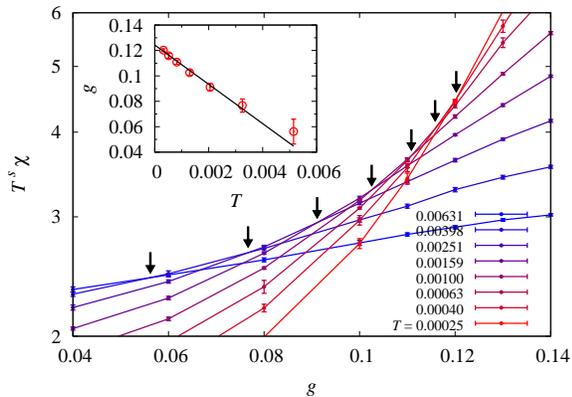}
	\end{center}
	\caption{(Color online) $T^s \chi$ as a function of $g$ for different values of $T$. From the intersection points, which are indicated by arrows, the critical point is determined. The inset shows an extrapolation of the intersection points to $T=0$.}
	\label{fig:cp}
\end{figure}

From the finite-temperature results in Figs.~\ref{fig:suscep-BFA} and \ref{fig:renorm}, 
two regimes have been identified: the Kondo regime for $g \lesssim 0.12$ and the local-moment regime for $g \gtrsim 0.12$. 
However, these data do not decide whether or not they are separated by a quantum critical point at $T=0$,
since we cannot exclude the possibility of finite but exponentially small Kondo temperature for $g \gtrsim 0.12$.
We need to extrapolate to lower temperatures in some way.
For this purpose,
we plot $T^{s} \chi$ as a function of $g$ for different temperatures in Fig.~\ref{fig:cp}.
The low-temperature expression for $\chi$ in Eq.~(\ref{eq:chi-BK}) 
indicates that $T^{s} \chi$ is independent of temperature provided $M_0=0$ (we define this point as $g=g_{\rm c}$), 
while $T^{s} \chi$ diverges due to $M_0 \neq 0$ for $g>g_{\rm c}$.
On the other hand, the local Fermi-liquid expression~(\ref{eq:chi-Kondo}) gives $T^{s} \chi=0$.
Hence, the intersection of lines for different temperatures in Fig.~\ref{fig:cp} gives an estimation of $g_{\rm c}$.
It turns out that the crossing point depends linearly on temperature down to $T=0.00025$ as shown in the inset of Fig.~\ref{fig:cp}.
From this result, we conclude a quantum phase transition between the local-moment state and the Kondo singlet state.
% By extrapolating it to $T=0$ with a linear line,}
% we obtain $g_{\rm c} \simeq 0.124$ with the error 0.001.
The critical coupling $g_{\rm c}$ is estimated at $g_{\rm c} \simeq 0.124$ by the linear extrapolation. 
% with the error 0.001.

\section{Summary}
\label{sec:summary}

We have developed an algorithm of CT-QMC for models including the spin-boson coupling, i.e., the Bose Kondo model and the Bose-Fermi Anderson model.
The algorithm covers up to three components for the bosonic field with XXZ-type anisotropy.
Simulations do not suffer from the sign problem, and therefore accurate computations can be achieved.
In this paper, we have restricted ourselves to $U=\infty$.
% in the simulation.
But, the formalism for the partition function, or the weight $W(\bm{\tau}, \bm{\mu})$, holds also for $U < \infty$,
so that only the update procedure should be modified to take account of the doubly occupied state.
One can also apply the present framework to the Kondo limit, i.e., the Bose-Fermi Kondo model.
For this purpose, the algorithm for the Kondo model (CT-J)\cite{Otsuki-CTQMC, Gull11} is available.

We have presented first numerical results for models with the SU(2) spin-boson coupling.
In the Bose Kondo model, 
we have observed the low-temperature static susceptibility 
consisting of the Curie term $T^{-1}$ as the leading term and a hidden bosonic fluctuating term $T^{-s}$, 
where $\rho_{\rm B}(\omega) \propto \omega^{s}$ with $s=0.2$.
This result demonstrates that the region $s=0.2$ does not belong to the critical phase, contradicting the perturbative RG approach which predicted the region $0<s<1$ as the critical phase, but being consistent with 
recent numerical calculations
for the XY-type coupling.\cite{Guo12}
Identifying the critical phase close to $s=1$ and determining the phase diagram require further careful computations, 
since the distinction between $T^{-1}$ and $T^{-s}$ becomes numerically harder as $s$ approaches 1.
This issue will be investigated elsewhere.

Including hybridization with the fermionic field, i.e., in the Bose-Fermi Anderson model,
we have investigated the evolution from the local-moment regime observed in the pure bosonic system to the Kondo regime, where the bosonic field is decoupled. 
By extrapolating some quantity to $T=0$, 
we have concluded
that these two states are separated by a quantum critical point,
at which the quasiparticle energy scale and the effective moment vanish from each side of $g$.
On the other hand, the energy scale of the bosonic fluctuations is not affected by the hybridization.
% As a result, the static susceptibility exhibits the power-law singularity $\chi \propto T^{-s}$ as the leading term at the critical point.
As a result, the power-law singularity $\chi \propto T^{-s}$ is expected as the leading term at the critical point
% For identification of the critical point,
in common with the critical point between the Kondo phase and the critical phase close to $s=1$.
% This exponent is the same as that for the critical point between the critical phase and the Kondo phase close $s=1$.\cite{}

The method presented in this paper can be applied to lattice models such as the Heisenberg model and the $t$-$J$ model by means of the extended DMFT.
Applications to lattice models as well as detailed investigations of the impurity models are left for future issues.

% \begin{acknowledgments}
We thank M. Vojta for useful comments on the manuscript.
The author is supported by JSPS Postdoctoral Fellowships for Research Abroad.
% \end{acknowledgments}


\begin{thebibliography}{99}

\bibitem{Rubtsov05} A.N. Rubtsov, V.V. Savkin and A.I. Lichtenstein, Phys. Rev. B \textbf{72}, 035122 (2005).
\bibitem{Gull11} For a review, see E. Gull, A. J. Millis, A. I. Lichtenstein, A. N. Rubtsov. M. Troyer, and P. Werner, Rev. Mod. Phys. \textbf{83}, 349 (2011).
\bibitem{Werner06} P. Werner, A. Comanac, L. de' Medici, M. Troyer, and A. J. Millis, Phys. Rev. Lett. \textbf{97}, 076405 (2006);
P. Werner and A. J. Millis, Phys. Rev. B \textbf{74}, 155107 (2006).
\bibitem{Haule07} K. Haule, Phys. Rev. B \textbf{75}, 155113 (2007).
\bibitem{Lauchli09} A. M. L\"auchli and P. Werner, Phys. Rev. B \textbf{80}, 235117 (2009).
\bibitem{Otsuki-CTQMC} J. Otsuki, H. Kusunose, P. Werner and Y. Kuramoto, J. Phys. Soc. Jpn. \textbf{76}, 114707 (2007).
\bibitem{Hoshino09} S. Hoshino, J. Otsuki, and Y. Kuramoto, J. Phys. Soc. Jpn. \textbf{78}, 074719 (2009).

\bibitem{Werner07} P. Werner and A. J. Millis, Phys. Rev. Lett. \textbf{99}, 146404 (2007);
P. Werner and A. J. Millis, Phys. Rev. Lett. \textbf{104}, 146401 (2010).
\bibitem{Pixley10} J. H. Pixley, S. Kirchner, M.T. Glossop, Q. Si, J. Phys.: Conf. Series \textbf{273}, 012050 (2011).

\bibitem{Bray-Moore80} A. J. Bray and M. A. Moore, J. Phys. C: Solid State Phys. \textbf{13}, L655 (1980).
\bibitem{Sachdev-Ye93} S. Sachdev and J. Ye, Phys. Rev. Lett. \textbf{70}, 3339 (1993).\bibitem{Grempel98} D. R. Grempel and M. J. Rozenberg, Phys. Rev. Lett. \textbf{80}, 389 (1998).
\bibitem{Georges00} A. Georges, O. Parcollet, and S. Sachdev, Phys. Rev. Lett. \textbf{85}, 840 (2000);
Phys. Rev. B \textbf{63}, 134406 (2001).

\bibitem{Kuramoto-Fukushima98} Y. Kuramoto and N. Fukushima, J. Phys. Soc. Jpn. \textbf{67}, 583 (1998);
N. Fukushima and Y. Kuramoto, J. Phys. Soc. Jpn. \textbf{67}, 2460 (1998).
\bibitem{Vojta00} M. Vojta, C. Buragohain, and S. Sachdev, Phys. Rev. B \textbf{61}, 15152 (2000).

\bibitem{Georges96} A. Georges, G. Kotliar, W. Krauth and M. J. Rozenberg, Rev. Mod. Phys. \textbf{68}, 13 (1996). 
\bibitem{Parcollet99} O. Parcollet and A. Georges, Phys. Rev. B \textbf{59}, 5341 (1999).

\bibitem{Smith-Si00} J. L. Smith and Q. Si, Phys. Rev. B \textbf{61}, 5184 (2000).
\bibitem{Haule02} K. Haule, A. Rosch, J. Kroha, and P. W\"olfle, Phys. Rev. Lett. \textbf{89}, 236402 (2002);
Phys. Rev. B \textbf{68}, 155119 (2003).
\bibitem{Sun-Kotliar02}
P. Sun and G. Kotliar, Phys. Rev. B \textbf{66}, 085120 (2002).

\bibitem{Rubtsov12} A. N. Rubtsov, M. I. Katsnelson, and A. I. Lichtenstein, Ann. Phys. \textbf{327}, 1320 (2012). 

\bibitem{Leggett87} A. J. Leggett, S. Chakravarty, A. T. Dorsey, A. Garg, and W. Zwerger, Rev. Mod. Phys. \textbf{59}, 1 (1987).
\bibitem{Bulla05} R. Bulla, H.-J. Lee, N.-H. Tong, and M. Vojta, Phys. Rev. B \textbf{71}, 045122 (2005).

% Bose-Fermi Kondo
\bibitem{Si01} Q. Si, S. Rabello, K. Ingersent, and J. L. Smith, Nature (London) \textbf{413}, 804 (2001)
\bibitem{Vojta06} For a review, see M. Vojta, Philos. Mag. \textbf{86}, 1807 (2006).
\bibitem{Sengupta00} A. M. Sengupta, Phys. Rev. B \textbf{61}, 4041 (2000).
\bibitem{Zhu-Si02} L. Zhu and Q. Si, Phys. Rev. B \textbf{66}, 024426 (2002).
\bibitem{Zarand-Demler02} G. Zar\'and and E. Demler, Phys. Rev. B \textbf{66}, 024427 (2002).
\bibitem{Glossop05} M. T. Glossop and K. Ingersent, Phys. Rev. Lett. \textbf{95}, 067202 (2005);
Phys. Rev. B \textbf{75}, 104410 (2007).
\bibitem{Winter09} A. Winter, H. Rieger, M. Vojta, and R. Bulla, Phys. Rev. Lett. \textbf{102}, 030601 (2009).
\bibitem{Guo12} 
C. Guo, A. Weichselbaum, J. von Delft, and M. Vojta, 
Phys. Rev. Lett. \textbf{108}, 160401 (2012).
\bibitem{note-NRG} It is known that a simple application of NRG to the spin-boson model may produce errors. See Ref.~\cite{Vojta12} for detail.
\bibitem{Vojta12} 
M. Vojta, 
Phys. Rev. B \textbf{85}, 115113 (2012).



\bibitem{footnote-pm}
Although the Hamiltonian (\ref{eq:H_imp}) leads to the symmetry
$\langle T_{\tau} \phi^{+}(\tau) \phi^{-} \rangle_0 = \langle T_{\tau} \phi^{-}(\tau) \phi^{+} \rangle_0$, we do not use it in the formulation.
Hence, all the expressions below are valid also for the case 
$\langle T_{\tau} \phi^{+}(\tau) \phi^{-} \rangle_0 \neq \langle T_{\tau} \phi^{-}(\tau) \phi^{+} \rangle_0$.

\bibitem{Anders11} P. Anders, E. Gull, L. Pollet, M. Troyer, and P. Werner, New J. Phys. \textbf{13}, 075013 (2011).
\bibitem{footnote-extrap} If we use the inverse of a quadratic function for the extrapolation, the convergence to a low-temperature value is faster, but the result is more sensitive to statistical errors except for low temperatures. The same is true of the Pad\'e approximation.
\bibitem{Hewson} See for example, A. C. Hewson, \textit{The Kondo problem to heavy fermions}  (Cambridge University Press, Cambridge, 1993).




\end{thebibliography}
\end{document}